\documentclass{PoS}

\usepackage{url}

\title{Performance of the Cherenkov Telescope Array }

\ShortTitle{Performance of the CTA}

\author{\speaker{G. Maier}$^{1}$, L. Arrabito$^{2}$, K. Bernl{\"o}hr$^{3}$, J. Bregeon$^{2}$, P. Cumani$^{4}$, T. Hassan$^{4}$, \mbox{A. Moralejo$^{4}$} for the CTA Consortium\footnote{Full consortium list at http://cta-observatory.org}\\
       $^{1}$ Deutsches Elektronen-Synchrotron (DESY), Platanenallee 6, D-15738 Zeuthen, Germany\\
       $^{2}$Laboratoire Univers et Particules de Montpellier - UMR5299, Universit\'e de Montpellier - CNRS/IN2P3, Place Eug\`ene Bataillon - CC 72, 34095 Montpellier C\'edex 05 France\\
       $^{3}$Max-Planck-Institut f{\"u}r Kernphysik, P.O. Box 103980, 
        D-69029 Heidelberg, Germany\\
        $^{4}$Institut de Fisica d'Altes Energies (IFAE), The Barcelona Institute of Science and Technology, Campus UAB, 08193 Bellaterra (Barcelona) Spain\\
       E-mail: \email{gernot.maier@desy.de}}

\abstract{The Cherenkov Telescope Array (CTA) will be the world's largest and by far most sensitive observatory for high-energy gamma rays. It will be capable of detecting gamma rays from extremely faint sources with unprecedented precision on energy and direction in the energy range from 20 GeV to more than 300 TeV. The performance of the future CTA observatory derived from detailed Monte Carlo simulations is presented in this contribution for the two CTA sites in Paranal (Chile) and on the La Palma island (Spain).}

\FullConference{35th International Cosmic Ray Conference\\
		 10-20 July, 2017\\
		 Bexco, Busan, Korea}

\begin{document}

\section{Introduction}

The Cherenkov Telescope Array (CTA) will be the next-generation gamma-ray observatory \cite{CTA,Ong-2017} and the world largest instrument for the observation of high-energy photons. 
It will consist of two arrays, each consisting of a large number of imaging atmospheric Cherenkov telescopes.
These telescopes observe the faint light emitted through the Cherenkov effect when the secondary particles produced in the cascade following the interaction of the high-energetic astrophysical gamma rays with the atmosphere pass through the air. 
Among the unique capabilities of the CTA Observatory are:
\begin{itemize}

\item A wide energy range covering an interval from 20 GeV to 300 TeV. This will enable the measurement of spectral energy distributions consistently over four decades of energies, both to detect high-redshift sources at low energies and to reach the energies of extreme accelerators in our Galaxy.

\item The large field of view of over 8 degrees of the CTA medium- and small sized-telescopes will allow to perform efficiently a deep survey of the complete Galactic plane and a survey of a significant part of the extragalactic sky.

\item A large effective area of $5\times 10^4$ m$^2$ at 50 GeV, $10^6$ m$^2$ at 1 TeV, and $5\times 10^6$ m$^2$  at 10 TeV, which will provide orders of magnitude better sensitivity than current instruments to short-term transient phenomena like GRBs or flaring active galactic nuclei \cite{Funk:2012}.

\item The extremely powerful suppression of events from background cosmic-ray nucleons, which results in an increase in sensitivity by a factor of five to ten as compared to the current instruments.

\item An angular resolution reaching two arcminutes, allowing to image extended sources in unprecedented detail.

\item An energy resolution and systematic uncertainty on the energy of well below 10\%, which provides sensitivity to features in the energy spectra (like e.g.~lines or cutoffs).

\item A view of the entire sky through the operation of arrays in both hemispheres.

\item A large number of different observation modes ranging from observation with the full arrays for highest sensitivity, a divergent-pointing mode providing a instantaneous field of view of 20 deg diameter \cite{Gerard:2015}, or the operation of  several sub arrays for the simultaneous observations of several targets.

\end{itemize}

The key features providing the huge improvement in capabilities of CTA relative to existing observatories are the deployment of a much larger number of telescopes, the deployment of three different telescope sizes to provide sensitivity across the large energy range (CTA will consist of large-sized, medium-sized, and small-sized telescopes), and the much larger field-of-view compared to operating observatories of up to 8 degrees.

\begin{table}[ht]
\centering
\begin{tabular}{c c c c c c c c}
\hline
Site & Longitude, Latitude & Altitude & B$_x$ & B$_z$ & LSTs & MSTs & SSTs   \\
 & [deg] & [m] & [$\mu$T]  & [$\mu$T]  \\
\hline
Paranal & 70.3W, 24.07S  & 2150 & 21.4 &  -8.9 & 4 & 25 & 70  \\
\hline
La Palma & 17.89W, 28.76N & 2180 & 30.8 & 23.2 & 4 & 15 & -   \\
\end{tabular}
\caption{\label{tab:sites} Site characteristics for the CTA Paranal and La Palma sites.
The horizontal and vertical component of the geomagnetic field are given by B$_x$ and B$_z$.
The number of telescopes of each type for each site are given for large-sized telescopes (LSTs), medium-sized telescopes (MSTs), and small-sized telescopes (SSTs).}
\end{table}

\begin{figure}
\centering\includegraphics[width=0.48\linewidth]{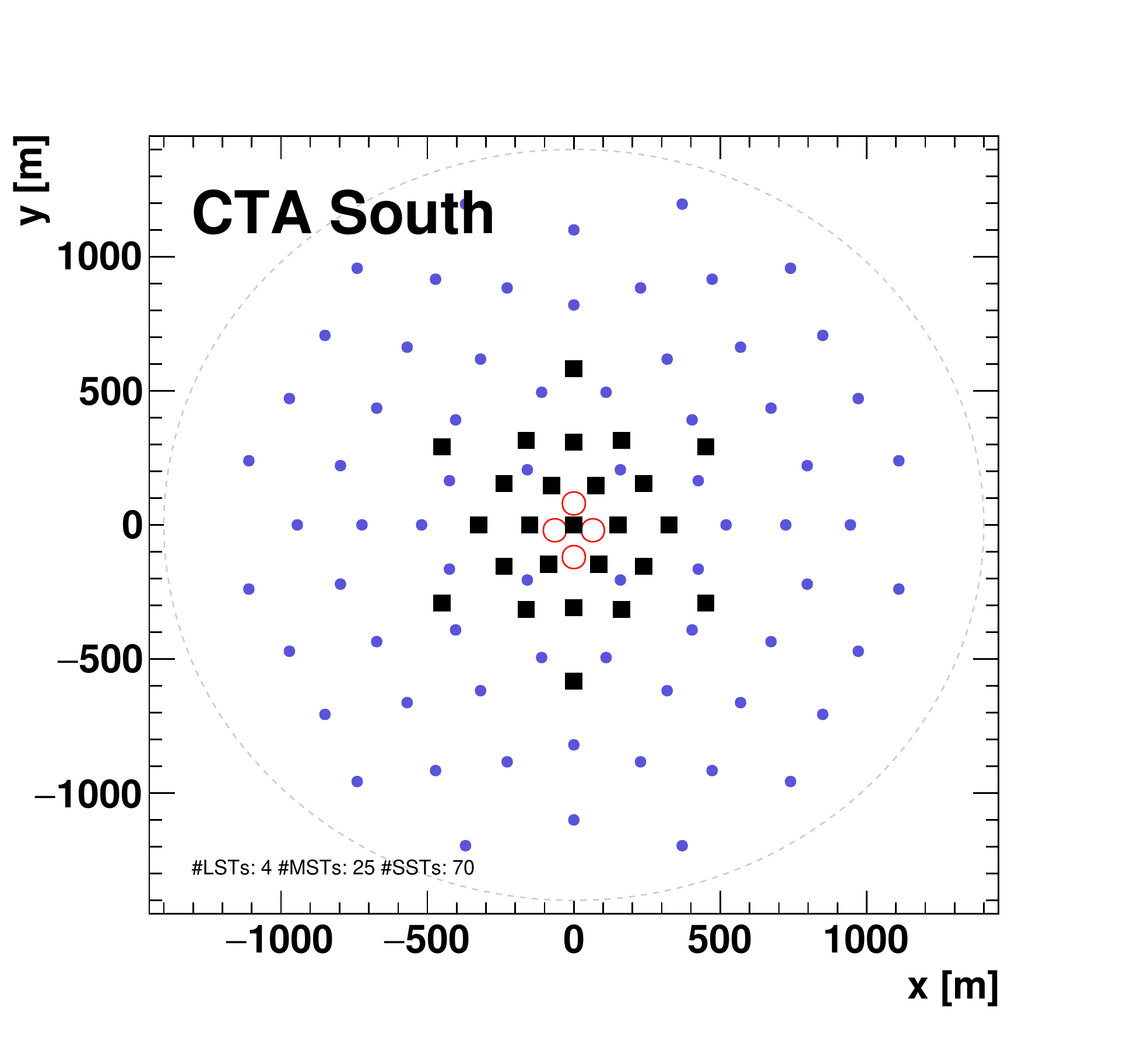}
\centering\includegraphics[width=0.48\linewidth]{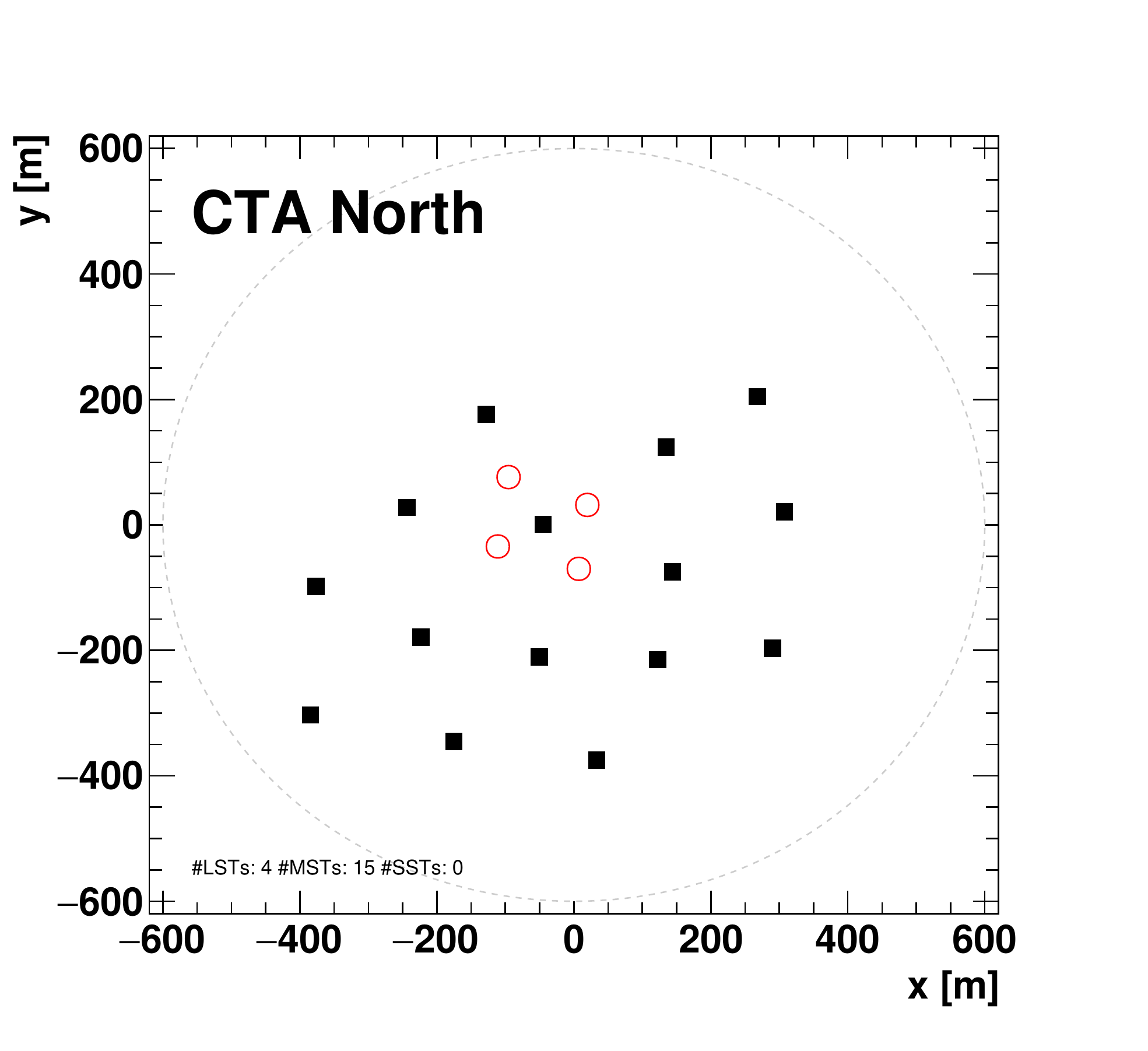}
\caption{\label{fig:array}
Telescope layouts for the Southern (left) and Northern CTA site (right).
The open circles indicate large-sized telescopes, the filled squares medium-sized telescope, and the filled points small-sized telescopes (southern site only).
}
\end{figure}

In this work, an overview of the performances for the CTA observatory is provided.
It is derived from detailed Monte Carlo simulations of the instrument.
CTA will consist of:
\begin{itemize}
\item A large array of 99 telescopes of three different sizes to be built at the Paranal site in Chile

\item An array consisting of 19 telescopes of two different types (large- and medium-sized telescopes) to be built on the Roque de los Muchachos, La Palma, Spain

\end{itemize}
Table \ref{tab:sites} gives an overview of the site characteristics and the number of telescopes of each type at the two CTA sites.
The exact arrangement of the telescopes on the ground are the result of an extensive and detailed optimisation procedure   \cite{Cumani:2017} using the same simulation setup as described in Section 2.
Figure \ref{fig:array} shows the layout for both CTA sites used for the determination of the performance of the CTA observatory.

\section{Monte Carlo simulations, reconstruction \& analysis}

The performance values for CTA presented in the following are derived from detailed Monte Carlo (MC) simulations of the CTA instrument based on the CORSIKA air shower code and the telescope simulation tool {\it sim\_telarray} \cite{Bernloehr-2008}. 
The MC simulations are similar to the one presented in \cite{Bernloehr-2013} and \cite{Hassan-2015}, but for an updated detector model of the CTA telescopes (so called {\em production 3(b)} or {\em prod3(b)}).

We assume to observe a gamma-ray source with a spectral shape following a power law with $E^{-2.6}$.
None of the results (e.g.~differential flux sensitivities, effective areas, angular or energy resolutions) as presented below depend noticeable on the assumed spectral shape of the gamma-ray source
(this is in contrast to integral sensitivity, which heavily depends on the assumed spectral index). 
Background cosmic-ray spectra of proton and electron/positron particle types are assumed using recent measurements for the spectra of both.
Heavier nuclei like cosmic-ray helium are not simulated, as studies show that CTA can almost completely suppress events from heavy nuclei (Z$>$2).

The optics, focal plane, and trigger simulations include all currently discussed telescopes and camera types of CTA \cite{CTA,Ong-2017}:
large-sized telescopes (LST) with a 23 m diameter optical dish, 12 m diameter medium-sized telescopes (MST) with two types of Cherenkov photon cameras (FlashCam or NectarCam), dual mirror medium-sized telescopes with a 9.6 m aperture (MS-SCT), and three types of small-sized telescopes (SSTs; 4 m dish diameter; ASTRI, GCT, SST-1m).
None of the results presented in these proceedings depend on the choice of the MST or SST type.

The official CTA reconstruction and analysis pipeline is currently under development, and is not available yet for 
sensitivity estimations.
Two reconstruction chains developed for the analysis of data from currently operating instruments have been adapted for the analysis of CTA MC simulations (MARS from MAGIC \cite{Moralejo-2009}; Eventdisplay from VERITAS \cite{Maier:2017}).
The reconstruction packages perform  analyses based on image parameterisations, with different choices of algorithms for image cleaning, background suppression (Boosted Decision Trees vs.~Random Forest) and energy reconstruction (look-up tables in combination with boosted decision trees vs.~Random Forest).
As both analyses are "classical" ones (based on parametrised shower images),  some improvement is expected with the use of more sophisticated techniques fully exploiting pixel-wise information, like e.g.~image templates methods \cite{LeBohec-1998, Parson-2014}.
All results presented here have been cross-checked with both analysis chains.

Nominal telescope pointing is assumed in the MC simulations, with all telescopes pointing directions parallel to one another 
(performance estimation for other pointing modes, e.g.~divergent pointing will be provided in the future).
Performance estimations are available for two zenith angles ($20^{\mathrm{o}}$ and $40^{\mathrm{o}}$), and for each zenith angle for 
two different azimuth angles (corresponding to pointing towards the magnetic North and South). 
Significant performance differences are found between the two azimuthal pointing directions, especially for the Northern site, as the impact of the geomagnetic field is large enough to influence notably the air shower development. 
The results presented here are obtained by averaging between the two azimuthal pointings

\section{Performance of CTA}

%
%
 \begin{figure}
 \centering
     \includegraphics[width=.999\textwidth]{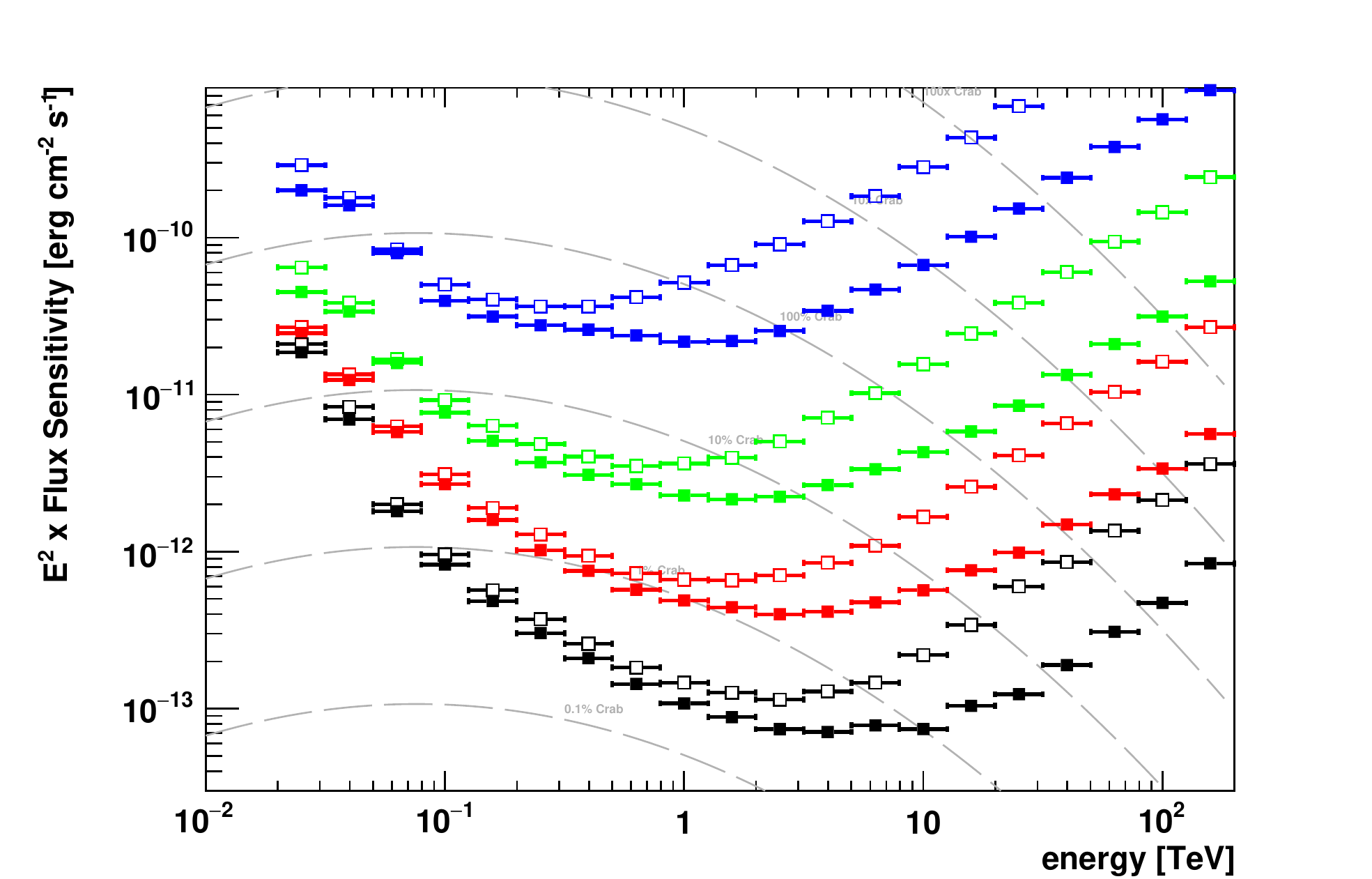}
     \caption{Differential energy flux sensitivities for CTA South (Paranal site; filled symbols) and CTA North (La Palma site; open symbols) for five standard deviation detections in five independent logarithmic bins per decade in energy and four different observation times (50 h: black symbols; 5 h: red symbols; 30 min: green symbols; 100 s: blue symbols).
    Additional criteria are applied to require at least ten detected gamma rays per energy bin and a signal/background ratio of at least 1/20.
    All flux sensitivities are calculated for a zenith angle of 20 degrees and average pointing directions.
    Horizontal lines indicate the width of the energy bin.
    }
     \label{fig1}
  \end{figure}

 \begin{figure}
 \centering
     \includegraphics[width=.49\textwidth]{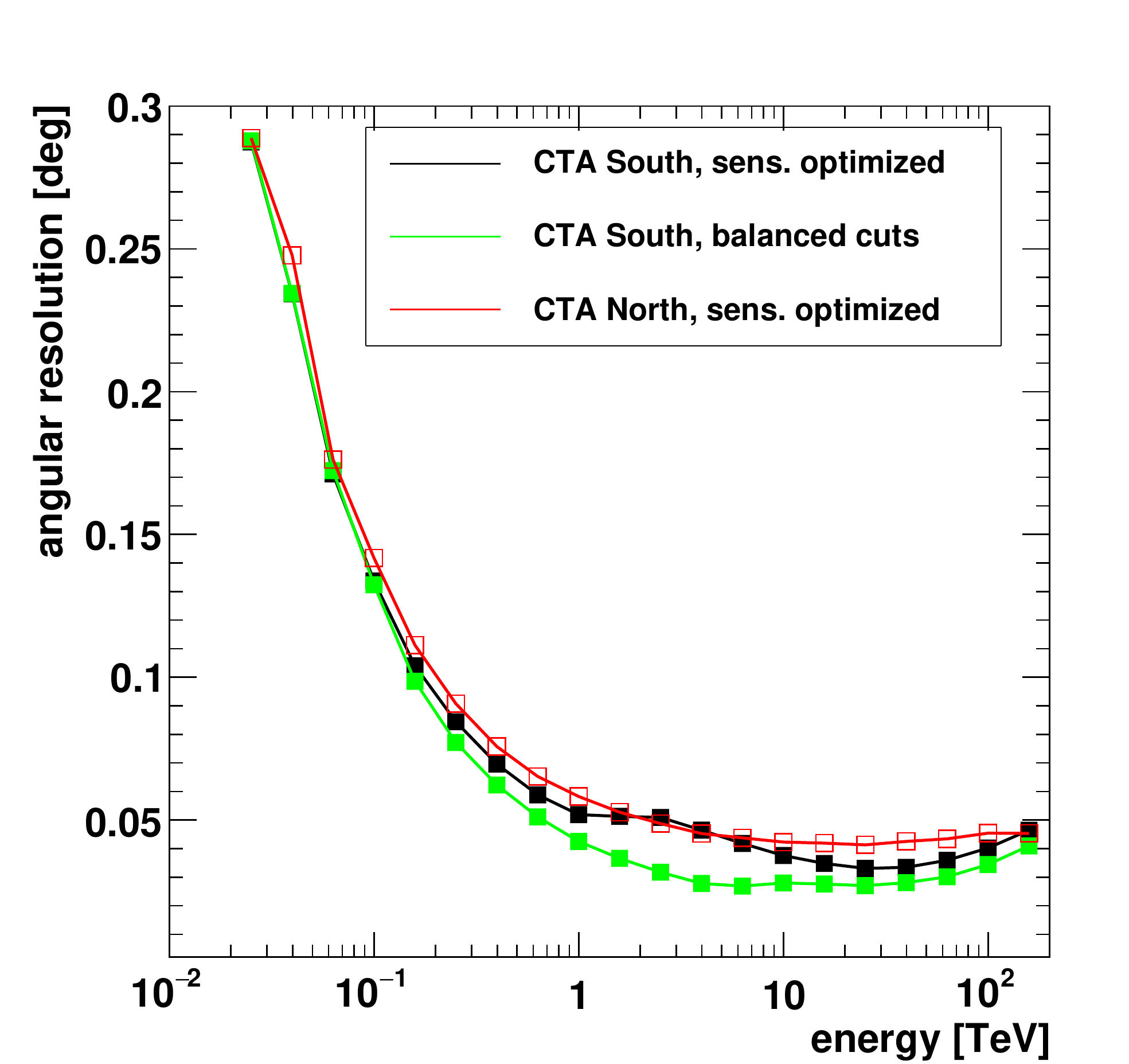}
      \includegraphics[width=.49\textwidth]{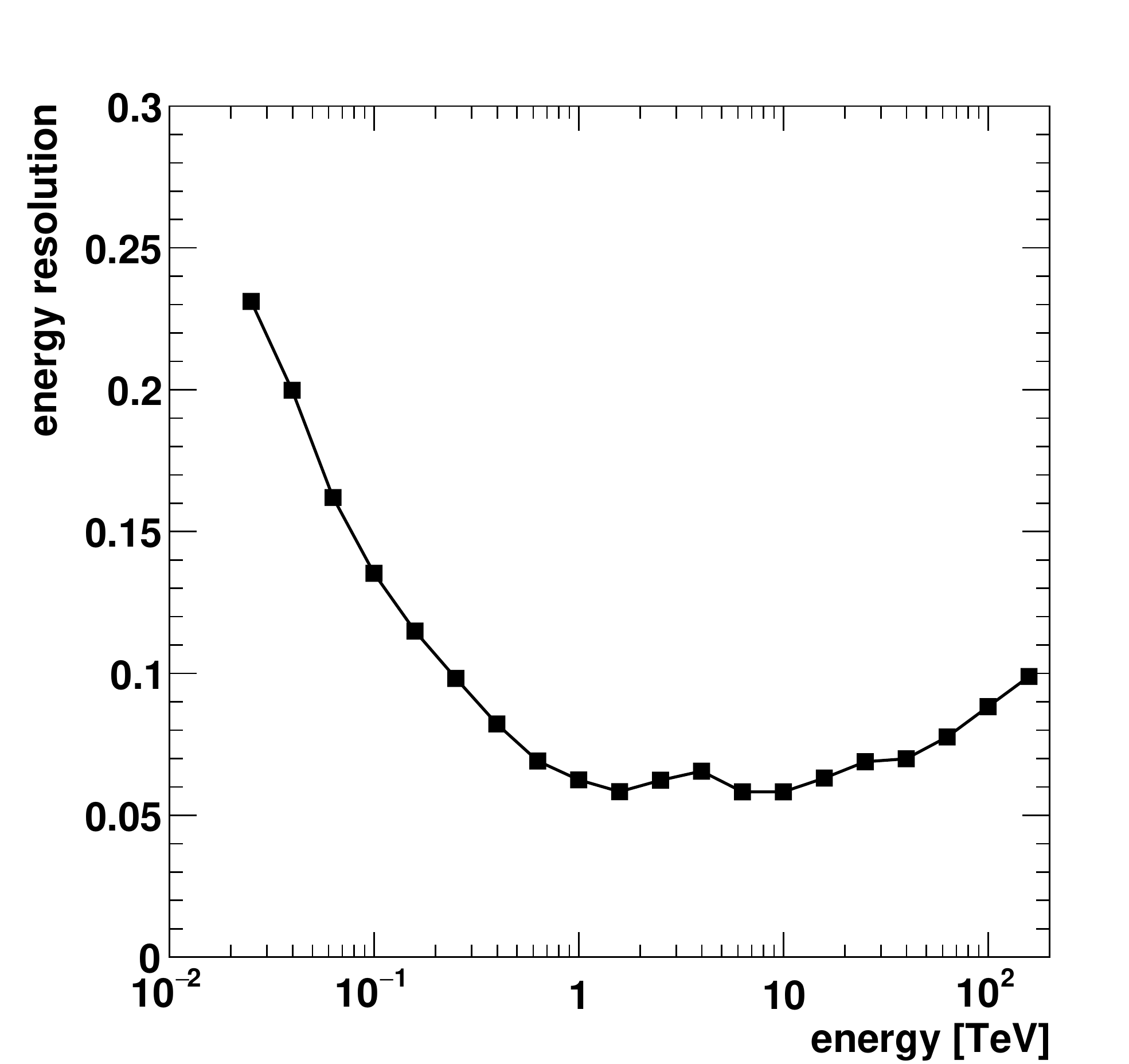}
     \caption{Left: The angular resolution vs. reconstructed energy curve shows the angle within which 68\% of reconstructed gamma-rays fall, relative to the true direction. 
     Angular resolution is evaluated with analysis cuts optimised for sensitivity (black and red curves corresponding to the differential sensitivity curves for 50h in Figure \ref{fig1}) and cuts optimised for a balance between sensitivity and angular resolution (green curve)
     Right: Energy resolution vs reconstructed energy.
     Energy resolution is defined such that 68\% of gamma rays will have true energy within $\Delta$E of their reconstructed energy.
     }
     \label{fig2}
  \end{figure}

 \begin{figure}
 \centering
     \includegraphics[width=.49\textwidth]{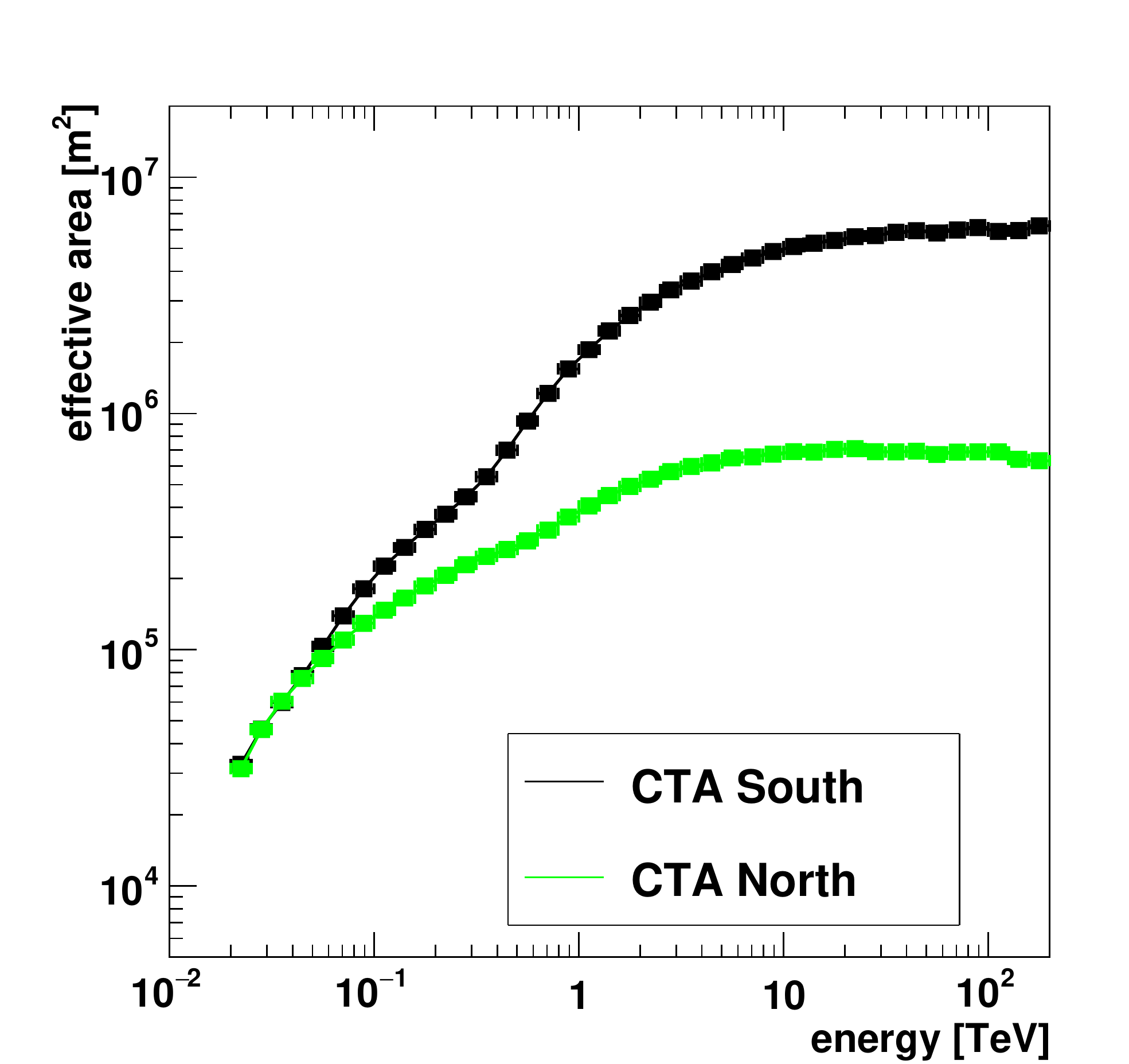}
      \includegraphics[width=.49\textwidth]{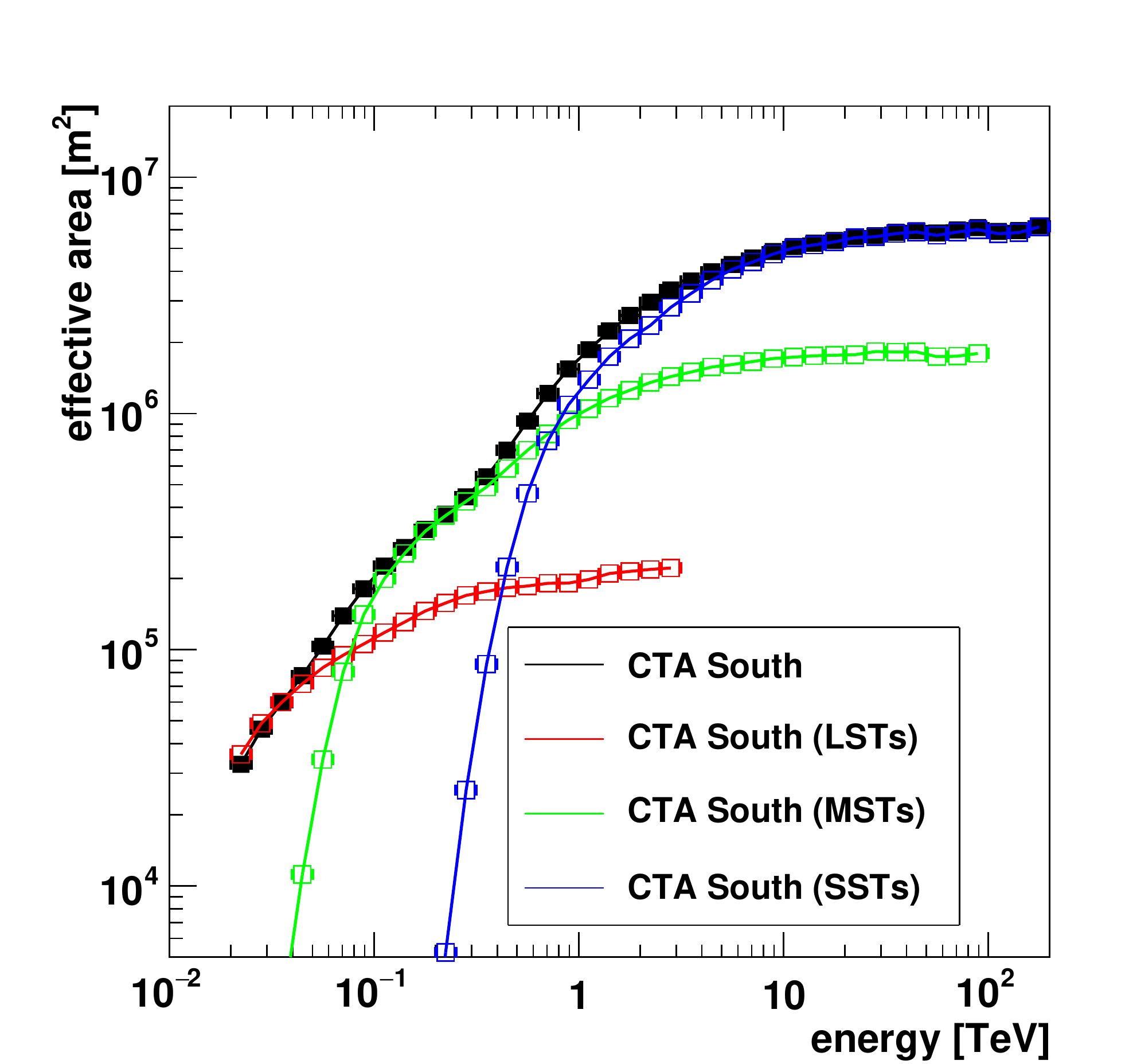}
     \caption{Effective collection area vs. true energy for gamma rays passing gamma-hadron separation cuts.
     Left: Effective collection area for CTA South and North.
     Right: Effective collection area for the CTA South array (see Figure \ref{fig:array} left), and the sub arrays of individual telescopes types (LSTs: large-sized telescopes; MSTs: medium-sized telescopes; SSTs: small-sized telescopes).  
     All curves correspond to the 30-min sensitivity curves given in Figure \ref{fig1}.}
     \label{fig3}
  \end{figure}

 \begin{figure}
 \centering
     \includegraphics[width=.99\textwidth]{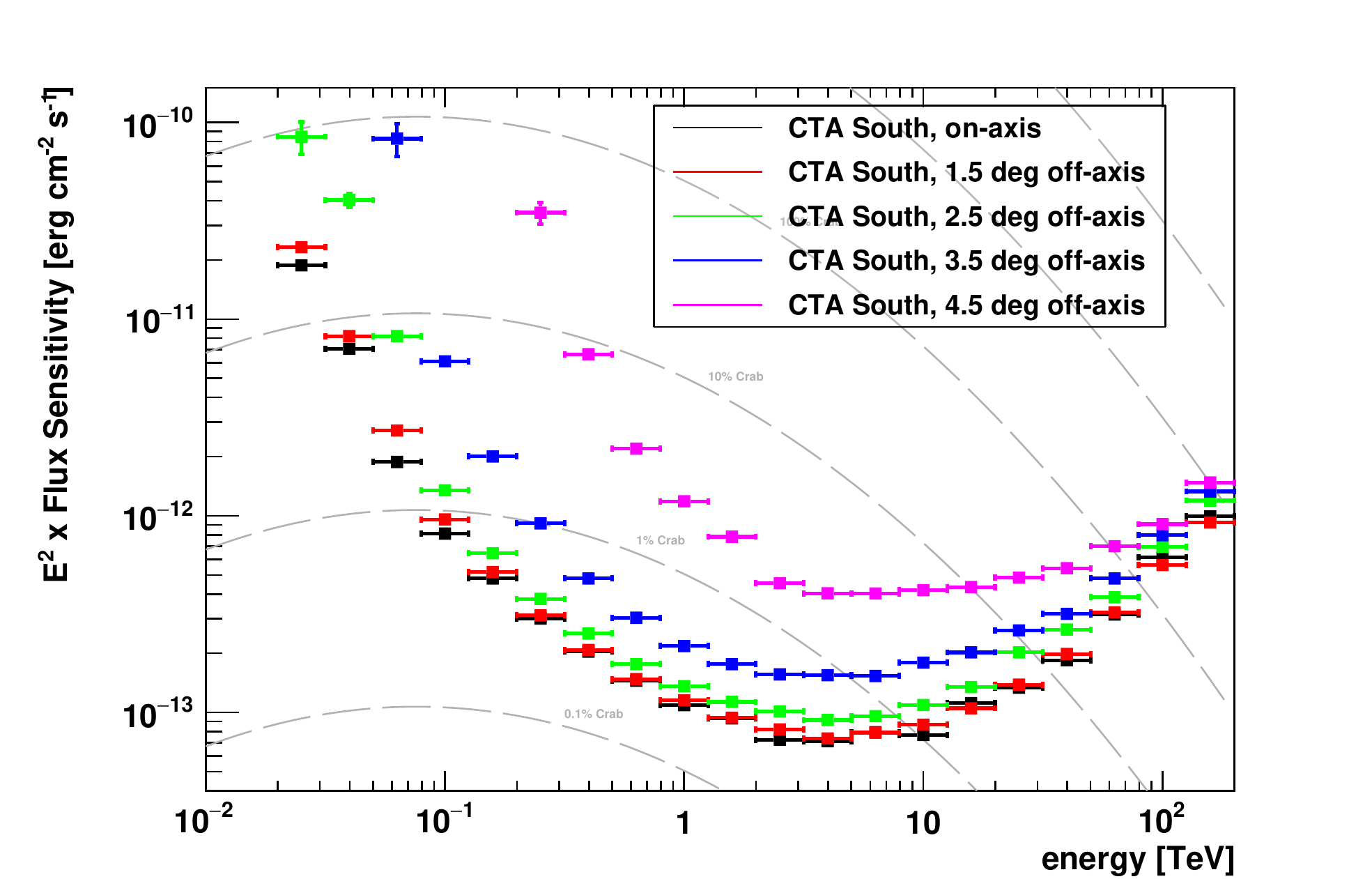}
     \caption{Differential energy flux sensitivities for CTA South (Paranal site) for a point source located at different distances (off-axis angles) to the camera centre.
     The assumed observation time is 50 h and the average zenith angle for these simulations is 20 degrees.
   For more details, see caption of Figure \ref{fig1}.  
    }
     \label{fig4}
  \end{figure}

Figures \ref{fig1}-\ref{fig4} give a broad overview of the sensitivity and performance of the CTA arrays for both sites.
The main performance metric for CTA is the differential energy flux sensitivity (Figure \ref{fig1}), given for the two CTA sites and analyses optimised for four different observation times.
Note that the applied definition for sensitivity requires a detection significance of $5\sigma$ per energy bin.
Additional criteria are applied to require at least ten detected gamma rays per energy bin and a signal/background ratio of at least 1/20.

The sensitivity at low energies ($<$200 GeV) will be very similar for both CTA arrays, achieving a high sensitivity down to energies of about 20 GeV.
The measurement is background limited in this energy regime, with the main contribution to the background from low-energy cosmic-ray protons with images looking similar to those of  gamma rays.
In the mid-energy range (200 GeV to several TeV), CTA will be the first instrument to suppress the background due to hadronic cosmic-rays so well, that almost all residual background is due to cosmic-ray electrons.
The achieved sensitivity reaches 0.2\% of the flux of the Crab Nebula at these energies, roughly one decade better than the existing ground-based instrument H.E.S.S., MAGIC, or VERITAS.
At the highest energies, the measurement is essentially background free and mostly limited by the very large but finite effective area of CTA.  

The energy dispersion and angular resolution (68\% containment radius) of CTA depend strongly on the number of telescopes contributing to the measurement of an event and therefore on the energy of the incoming gamma ray.
Typical values for the energy resolution are 5-10\% and  0.1 degrees at 100 GeV and 0.02-0.03 degrees above 1 TeV for the angular resolution.
The angular and energy resolutions as shown in Figure \ref{fig2} represents very likely an underestimation of CTA capabilities: the analysis results shown are not optimised for best angular/energy resolution but for best sensitivity and the applied analysis method is (as mentioned earlier) not the most advanced one known in the field.
Improvements compared to the presented values are expected at energies below 1 TeV and especially for events with low image multiplicity.

The large effective area of CTA is critical for both the sensitivity to short transients and the characterization of sources beyond several hundreds of TeV.
Figure \ref{fig3} shows the comparison of the effective areas for the two CTA sites, and a comparison of the response of subarrays of same telescope types.
The latter clearly indicates the energies of highest sensitivity for the different CTA telescope types: 20 GeV to 200 GeV for the LSTs, 100 GeV to 5 TeV for the MSTs, and the highest energies for the SSTs.

The large field-of-view of up to 8 degrees for MSTs and SSTs results in a relatively small decrease of sensitivity as function of off-axis angle for CTA (Figure \ref{fig4}): the sensitivity is roughly constant up to a distance to the camera centre of 2 degrees for energies around 1 TeV. 
It reaches 50\% of the on-axis sensitivity at 3.7 degrees off-axis for similar energies.
Given its excellent off-axis performance, CTA will have a sensitivity at the edge of its field-of-view equal to the on-axis sensitivity of current instruments (for similar observation time).

\section{Conclusions}

The  simulation efforts in CTA have played a critical role in the planning, construction and completion of the instrument:  the setting of the baseline design with several telescope types distributed over a large area \cite{Bernloehr-2013}, the contribution to the CTA site selection process \cite{Hassan:2017}, the optimisation of the array layouts \cite{Cumani:2017}, the ongoing optimisation and verification of details of the instrument design, trigger, readout and reconstruction software, and soon the verification of the first data from on-site telescope prototypes.
The understanding of the CTA sensitivity under different observing conditions (e.g.~large zenith angle, high night-sky background or small subarrays) are among the next steps for the efforts of the CTA Monte Carlo group. 
The comprehensive characterisation of the CTA observatory by means of MC simulations presented in this work, including differential sensitivities, angular and energy resolution, effective collection areas and background rates will be made available through the public webpage of CTA\footnote{\url{https://www.cta-observatory.org/science/cta-performance/}} in electronic form.

\section*{Acknowledgments}

\noindent This work was conducted in the context of the CTA Consortium. 

\noindent We gratefully acknowledge financial support from the agencies and organizations listed here: 
\url{http://www.cta-observatory.org/consortium_acknowledgments}.

\end{document}